\documentclass[a4paper,12pt]{article}

\usepackage{amsmath}
\usepackage{amssymb}
\usepackage{amsfonts}
\usepackage{amsthm}
\usepackage{color}
\usepackage{cite}
\newcommand{\del}{\partial}

 \makeatletter
 \@addtoreset{equation}{section}
 \makeatother

\date{empty}
\begin{document}
\begin{titlepage}
\null
\begin{flushright}
November, 2020
\end{flushright}
\vskip 2cm
\begin{center}
{\large \bf 
Higher Derivative 
Supersymmetric
Nonlinear Sigma Models \\
on Hermitian Symmetric Spaces,
and BPS States Therein
}
\vskip 0.7cm
\normalsize
\renewcommand\thefootnote{\alph{footnote}}

{\large
Muneto Nitta${}^{\dagger}$\footnote{nitta(at)phys-h.keio.ac.jp}
and
Shin Sasaki${}^{\ddagger}$\footnote{shin-s(at)kitasato-u.ac.jp}
}
\vskip 0.5cm
  {\it
  ${}^{\dagger}$
  Department of Physics, and Research and Education Center for Natural
  Sciences, \\ 
  Keio University, Hiyoshi 4-1-1, Yokohama, Kanagawa 223-8521, Japan \\
  \vspace{0.2cm}
  ${}^{\ddagger}$
  Department of Physics,  Kitasato University \\
  Sagamihara 252-0373, Japan
  }
\vskip 0.7cm
\begin{abstract}
We formulate four-dimensional $\mathcal{N} = 1$ 
supersymmetric nonlinear sigma models
 on Hermitian symmetric spaces 
 with higher derivative terms, 
free from the auxiliary field problem 
and the Ostrogradski's ghosts, 
as gauged linear sigma models. 
We then study Bogomol'nyi-Prasad-Sommerfield equations preserving 
1/2 or 1/4 supersymmetries. 
We find that there are distinct branches, that we call 
canonical ($F=0$) and non-canonical ($F\neq 0$) branches, 
associated with solutions to auxiliary fields $F$ in chiral multiplets.
For the ${\mathbb C}P^N$ model,
we obtain a supersymmetric ${\mathbb C}P^N$ Skyrme-Faddeev model 
in the canonical branch while 
in the non-canonical branch 
the Lagrangian consists of solely
the ${\mathbb C}P^N$ Skyrme-Faddeev 
term without a canonical kinetic term.
These structures can be extended to the Grassmann manifold 
$G_{M,N} = SU(M)/[SU(M-N)\times SU(N) \times U(1)]$.
For  other Hermitian symmetric spaces such as 
the quadric surface  $Q^{N-2}=SO(N)/[SO(N-2) \times U(1)])$,
 we impose F-term (holomorphic) constraints for embedding  them 
 into ${\mathbb C}P^{N-1}$ or Grassmann manifold.
We find that these constraints are 
consistent in the canonical branch 
 but yield additional constraints on the dynamical fields 
  thus reducing the target spaces 
 in the non-canonical branch.
\end{abstract}
\end{center}

\end{titlepage}

\newpage
\setcounter{footnote}{0}
\renewcommand\thefootnote{\arabic{footnote}}
\pagenumbering{arabic}
\tableofcontents

\section{Introduction} \label{sect:introduction}

Nonlinear sigma models are typical examples of low-energy effective
theories.
When a global symmetry $G$ is spontaneously  broken
down to its subgroup $H$, there appear massless 
Nambu-Goldstone bosons dominant at low energy,
and their low energy dynamics can be
described by nonlinear sigma model whose target space is 
a coset space $G/H$ 
\cite{Coleman:1969sm,Callan:1969sn}.
If one wants to study physics at higher energy, 
 one needs  higher derivative correction terms,
typically appearing in the chiral perturbation theory 
of QCD \cite{Leutwyler:1993iq}.
Other examples of higher derivative terms 
can be found in various contexts such as 
the Skyrme model, 
low-energy effective theories of superstring theory, 
and 
worldvolume theories of topological solitons and branes 
such as Nambu-Goto and Dirac-Born-Infeld actions.

In supersymmetric theories, 
higher derivative terms often bring us troubles. 
It is known that constructing derivative terms
in the form of  $\del_{m} \Phi$ suffers from a technical issue called
the auxiliary field problem 
\cite{Gates:1995fx,Gates:2000rp}. 
This stems from the fact that the equation of
motion for the auxiliary field $F$ in the chiral superfield $\Phi$ ceases to
be an algebraic equation and becomes a kinematical one.
As a consequence, it is hard to integrate out the auxiliary field and
the interactions of physical fields are not apparent.
Supersymmetric extensions of the Wess-Zumino-Witten term 
\cite{Nemeschansky:1984cd} and the Skyrme-Faddeev model \cite{BeNeSc,Fr} 
are such examples with the auxiliary field problem, while 
low-energy Lagrangians for supersymmetric gauge theories 
in Refs.~\cite{Buchbinder:1994iw,Buchbinder:1994xq,Kuzenko:2014ypa} 
(see also Refs.~\cite{Banin:2006db,Gomes:2009ev,Gama:2011ws}),
and supersymmetric extensions of Dirac-Born-Infeld action 
\cite{RoTs,SaYaYo} 
and K-field theories
\cite{AdQuSaGuWe3,AdQuSaGuWe2} 
are free from this problem. 
A broad class of supersymmetric derivative terms 
free from the auxiliary field problem 
and 
the Ostrogradski's ghost \cite{Ostrogradski} 
was found as four-dimensional
$\mathcal{N} = 1$ supersymmetric higher derivative
chiral models 
formulated in terms of superfields 
\cite{Khoury:2010gb,KoLeOv2,KhLeOv2,KoLeOv,
Adam:2011hj,AdQuSaGuWe,
FaKe,Nitta:2014pwa,Nitta:2014fca,Nitta:2015uba,
Gudnason:2015ryh,Queiruga:2016yzd,Queiruga:2017blc}.\footnote{
Another possibility is to gauge them away 
\cite{Fujimori:2016udq} if there are Ostrogradski's higher derivative ghosts 
\cite{Antoniadis:2007xc,Dudas:2015vka}.
}
The model consists of a K\"ahler potential $K$ and a superpotential $W$
together with a $(2,2)$ K\"ahler tensor $\Lambda$
that
determines derivative corrections. 
It was applied to a new mechanism of supersymmetry breaking in modulated vacua 
 \cite{Nitta:2017yuf} (see also Refs.~\cite{Nitta:2017mgk}).
The general formulation was extended for other superfields:
the most general ghost-free (and tachyon-free) 
higher derivative ${\cal N}=1$ supersymmetric 
Lagrangian for vector multiplets (1-form gauge fields)  \cite{Fujimori:2017kyi}
and 3-form gauge field \cite{Nitta:2018yzb}.
As for nonlinear sigma models of chiral superfields, 
only known examples are the supersymmetric 
${\mathbb C}P^1$ model (or baby Skyrme model)  
\cite{Adam:2011hj,AdQuSaGuWe,Nitta:2014pwa,Bolognesi:2014ova,Queiruga:2016jqu,Gudnason:2020tps} 
and the supersymmetric Skyrme model \cite{Gudnason:2015ryh}.
When one solves the auxiliary field equations of motion 
for the auxiliary field $F$, it in general 
allows more than one solutions $F=0$ and $F \neq 0$.
The former is called the canonical branch and the latter the non-canonical branch. 
The bosonic part of the Lagrangian of the canonical branch consists of 
a canonical kinetic term and a four derivative term 
for the ${\mathbb C}P^1$ model \cite{Faddeev:1996zj}.
This admits lump (sigma model instanton) solutions identical to 
those in the ${\mathbb C}P^1$ model without a higher derivative term 
\cite{Polyakov:1975yp},
as Bogomol'nyi-Prasad-Sommerfield (BPS) states  
preserving 1/2 supersymmetric charges 
among the original supersymmetry 
\cite{Nitta:2014pwa}. 
On the other hand, in the non-canonical branch, the  bosonic part of the Lagrangian 
consists of only 
the four derivative term 
of the Skyrme-Faddeev type \cite{Faddeev:1996zj} 
(that is without fourth order time derivatives)
without any canonical kinetic term.
This  admits a compact baby Skyrmion \cite{Adam:2009px} 
which is a BPS state  
preserving 1/4 supersymmetry \cite{Nitta:2014pwa}. 
While the general Lagrangian in superspace \cite{Khoury:2010gb} 
can be defined on any K\"ahler target spaces,
solving auxiliary field equations consisting of more than one chiral multiplets 
is still an open problem,
 except for a single matrix chiral superfield 
 \cite{Gudnason:2015ryh}.
 
The purpose of this paper is to present 
higher derivative supersymmetric nonlinear sigma models 
with a wider class of target spaces -- Hermitian symmetric spaces:
\begin{eqnarray}
\mathbb{C}P^{N-1} &=& SU(N)/[SU(N-1) \times U(1)],\nonumber \\ \nonumber
G_{M,N} &=& U(M)/[U(M-N) \times U(N)], \\ \nonumber
Q^{N-2} &=& SO(N)/[SO(N-2)\times U(1)], \\ \nonumber
&&SO(2N)/U(N), \quad\quad\quad
Sp(N)/U(N) , \\ 
&&E_6/[SO(10) \times U(1)] , \quad
E_7/[E_6 \times U(1)],
\end{eqnarray}
for which we solve auxiliary field equations. 
In the case without higher derivative terms, 
supersymmetric nonlinear sigma models on Hermitian symmetric spaces 
can be constructed by imposing supersymmetric constraints on gauged
linear sigma models \cite{Nitta:2014pwa}. 
This formulation is found to help us to solve auxiliary field equations
even with higher derivative terms. 
We employ the gauged linear sigma models 
with higher derivative terms for chiral multiplets \cite{Nitta:2015uba} 
and then take the strong gauge coupling (nonlinear sigma model) limit.
Solving auxiliary field equations 
yields the canonical and non-canonical branches for the on-shell actions. 
For the ${\mathbb C}P^N$ model, 
we obtain a supersymmetric extension of the ${\mathbb C}P^N$ Skyrme-Faddeev model  
\cite{Ferreira:2010jb,Ferreira:2011jy,Ferreira:2011ja,Eto:2012qda}
in the canonical branch,  
while the on-shell Lagrangian in the non-canonical  branch
consists solely of  the  ${\mathbb C}P^N$ Skyrme-Faddeev term 
(the term without fourth order time derivatives). 
We then study BPS states in these models, 
and find that lumps (sigma model instantons) identical to 
those without higher derivative terms 
remain 1/2 BPS states in the canonical branch, while
compact baby Skyrmions are 1/4 BPS states in the non-canonical branch.

The organization of this paper is as follows.
In the next section, we introduce the supersymmetric higher derivative term 
in the gauged chiral models. We show that a derivative term determined by a 
$(2,2)$ K\"ahler tensor provides supersymmetric higher derivative terms 
free from the auxiliary field problem.
We will see that there are two on-shell branches associated with the
solutions to the auxiliary field in the chiral multiplets.
In Section \ref{sect:NLSM}, we study the sigma model limit
of the gauged chiral models. 
We first focus on the sigma models whose target spaces are the
Hermitian symmetric spaces $\mathbb{C}P^{N-1}$ and $G_{M,N}$.
They are defined only by the D-term constraints.
We examine the limit in the two distinct
branches and write down the higher derivative terms 
in the nonlinear sigma models.
In Section \ref{sect:F-term}, we discuss the nonlinear sigma models 
defined by the F-term constraints in addition to D-term constraints,
which include the target spaces $Q^{N-2}$,
$SO(2N)/U(N)$, $Sp(N)/U(N)$, $E_6/[SO(10) \times U(1)]$, $E_7/[E_6
\times U(1)]$. 
We will show that the F-term constraints give additional
conditions on the target spaces in the non-canonical branch
thus reducing the target spaces, while they do not in the canonical branch.
This distinguishes the situation from the $\mathbb{C}P^{N-1}$ and
$G_{M,N}$ cases.

In Section \ref{sect:BPS}, we discuss the BPS states in the nonlinear
sigma models, 
and in Section \ref{sect:fermion}, we mention fermionic terms in the models.
Section \ref{sect:conclusion} is devoted to conclusion and
discussions.
As for a superfield notation, we follow the Wess-Bagger convention \cite{Wess:1992cp}.

\section{Supersymmetric higher derivative terms in the chiral model} \label{sect:SUSY_derivative_corrections}
In this section, we briefly introduce the supersymmetric higher 
derivative term to the chiral model that are free from the auxiliary field problem.
The Lagrangian is given by \cite{Khoury:2010gb, Nitta:2015uba}
\begin{align}
\mathcal{L} =& \  \int \! d^4 \theta \ K(\Phi,\Phi^{\dagger}) + 
\frac{1}{16} \int \! d^4 \theta \ \Lambda_{i\bar{j}k\bar{l}, ab} {}^{cd}
 (\Phi,\Phi^{\dagger}) 
D^{\alpha} \Phi^{ia} D_{\alpha} \Phi^{kb} \bar{D}_{\dot{\alpha}}
 \Phi^{\dagger \bar{j}}_c \bar{D}^{\dot{\alpha}} \Phi^{\dagger \bar{l}}_d
\notag \\
& \ 
+
\left(
\int \! d^2 \theta \ W (\Phi) + \mathrm{h.c.}
\right),
\label{eq:Lagrangian}
\end{align}
where $K (\Phi,\Phi^{\dagger})$ is a K\"ahler potential, $W(\Phi)$ is a
superpotential and $\Lambda_{i\bar{j}k\bar{l}}(\Phi,\Phi^\dagger)$ is 
a $(2,2)$ K\"ahler tensor defined by 
the chiral superfields $\Phi, \Phi^{\dagger}$
and their derivatives 
$D_{\alpha} \Phi$, $\bar{D}_{\dot{\alpha}} \Phi^{\dagger}$,
$\partial_m \Phi$, and $\partial_m \Phi^{\dagger}$.
We assume that the chiral superfields $\Phi^{ia} \ (i = 1, \ldots,
\mathrm{dim} \, G, a = 1, \ldots, \mathrm{dim} \, G')$ belong to 
the fundamental representations of 
the global and gauge symmetries $G$ and $G'$, respectively.
The four-dimensional $\mathcal{N} = 1$ chiral superfield is
expanded, in the chiral basis $y^m = x^m + i \theta \sigma^m
\bar{\theta}$, as 
\begin{align}
\Phi^{ia} = \varphi^{ia} (y) + \theta \psi^{ia} (y) + \theta^2 F^{ia} (y).
\end{align}
We stress that the pure bosonic component in the 
fourth derivative part 
\begin{align}
D^{\alpha} \Phi^{i a} D_{\alpha} \Phi^{kb} \bar{D}_{\dot{\alpha}}
 \Phi^{\dagger \bar{j}}_c \bar{D}^{\dot{\alpha}} \Phi^{\dagger \bar{l}}_d
\label{eq:4th_derivative}
\end{align}
saturates the Grassmann coordinates, and only the bosonic fields in the K\"ahler tensor $\Lambda_{i \bar{j} k \bar{l}}$
contribute to this pure bosonic term \cite{Khoury:2010gb}.
Indeed, the component expansion of the term \eqref{eq:4th_derivative} can be evaluated as 
\begin{align}
D^{\alpha} \Phi^{i\alpha} D_{\alpha} \Phi^{kb} \bar{D}_{\dot{\alpha}}
 \Phi^{\dagger \bar{j}}_c \bar{D}^{\dot{\alpha}} \Phi^{\dagger \bar{l}}_d
=& \
 16 \theta^2 \bar{\theta}^2
\left[
(\partial_m \varphi^{ia} \partial^m \varphi^{kb})(\partial_n \bar{\varphi}^{\bar{j}}_c \partial^n \bar{\varphi}^{\bar{l}}_d)
\right.
\notag \\
& \ 
\left.
- \frac{1}{2} 
\left(
\partial_m \varphi^{ia} F^{kb} + F^{ia} \partial_m \varphi^{kb}
\right)
\left(
\partial^m \bar{\varphi}^{\bar{j}}_c \bar{F}^{\bar{l}}_d + \bar{F}^{\bar{j}}_c \partial^m \bar{\varphi}^{\bar{l}}_d
\right)
+
F^{ia} \bar{F}^{\bar{j}}_c F^{kb} \bar{F}^{\bar{l}}_d
\right]
\notag \\
& \ + \cdots.
\end{align}
Here $\cdots$ are terms involving fermions.
Due to this remarkable structure, a large class of supersymmetric models with derivative corrections
can be realized by the Lagrangian \eqref{eq:Lagrangian}.
For example, the $\mathcal{N} = 1$ supersymmetric Dirac-Born-Infeld
model is given by the K\"ahler tensor (scalar) \cite{RoTs,SaYaYo}
\begin{align}
\Lambda (\Phi, \Phi^{\dagger}) = \frac{1}{1 + A + \sqrt{(1+A)^2 -B}},
 \qquad A = \del_m \Phi \del^m \Phi^{\dagger}, \qquad B = \del_m \Phi
 \del^m \Phi \del_n \Phi^{\dagger} \del^n \Phi^{\dagger}.
\end{align}
A supersymmetric Skyrme-Faddeev model is given by the K\"ahler metric $K_{\varphi
\bar{\varphi}} (1 + |\varphi|^2)^{-2}$ and the K\"ahler scalar \cite{Nitta:2014pwa}
\begin{align}
\Lambda (\Phi, \Phi^{\dagger}) = (\del_m \Phi \del^m \Phi \del_n \Phi^{\dagger} \del^n
 \Phi^{\dagger})^{-1} \frac{1}{(1 + \Phi \Phi^{\dagger})^4} 
\left[
(\del_m \Phi^{\dagger} \del^m \Phi)^2 - \del_m \Phi \del^m \Phi \del_n
 \Phi^{\dagger} \del^n \Phi^{\dagger}
\right].
\end{align}
The other examples include the Galileon inflation and the ghost
condensation models \cite{KoLeOv2, KhLeOv2} and a superconformal higher
derivative nonlinear sigma model \cite{Kuzenko:2020jzb}.

Now we make the global symmetry $G'$ be gauged by introducing 
the $\mathcal{N} = 1$ vector multiplet $V = V^{\hat{a}} 
T^{\hat{a}} \ (\hat{a} = 0, 1, \ldots, \mathrm{dim} \, \mathcal{G}')$:
\begin{align}
V = - (\theta \sigma^m \bar{\theta}) A_m (x) + i \theta^2 \bar{\theta}
 \bar{\lambda} (x) - i \bar{\theta}^2 \theta \lambda (x) + \frac{1}{2}
 \theta^2 \bar{\theta}^2 D(x).
\end{align}
Here we have employed the Wess-Zumino gauge and 
$T^{\hat{a}}$ are the generators of $\mathcal{G}'$ satisfying the normalization $\mathrm{Tr}
(T^{\hat{a}} T^{\hat{b}}) = k \delta^{\hat{a}\hat{b}}$.
In the derivative corrections, the vector multiplet is introduced in the gauge covariantization of the supercovariant derivative:
\begin{align}
D_{\alpha} \Phi^{ia} \to \mathcal{D}_{\alpha} \Phi^{ia} 
=
D_{\alpha} \Phi^{ia}
+
 (\Gamma_{\alpha})^a {}_b \Phi^{ib}.
\end{align}
The gauge superconnection $\Gamma_{\alpha}$ is defined by 
\begin{align}
(\Gamma_{\alpha})^a {}_b = e^{-2gV} D_{\alpha} e^{2gV},
\end{align}
where $g$ is the gauge coupling constant. 
The gauge invariant derivative term is given by a $(2,2)$ K\"ahler tensor of the following structure:
\begin{align}
\Lambda_{i\bar{j}k\bar{l},ab} {}^{cd} = 
\Lambda_{i\bar{j}k\bar{l}} (\Phi, \Phi^{\dagger})
 (e^{2gV})^c {}_a (e^{2gV})^d {}_b,
\end{align}
where $\Lambda_{i\bar{j}k\bar{l}}$ is composed of gauge invariant
quantities made of $\Phi, \Phi^{\dagger}, V$. 
These include the gauge covariant derivatives of the chiral superfields
$\mathcal{D}_{\alpha} \Phi$, $\bar{\mathcal{D}}_{\dot{\alpha}} \Phi^{\dagger}$.
We note that terms made of $\del_m \Phi$, $\del_m \Phi^{\dagger}$, which
are allowed in the ungauged case, are
forbidden in the gauged models due to supersymmetry and the gauge covariance.
Then, the gauged Lagrangian can be written as
\begin{align}
\mathcal{L} =& \ \int \! d^4 \theta \ K(\Phi,\Phi^{\dagger},V) -
 \frac{1}{16}
\int \! d^4 \theta \ \Lambda_{i\bar{j}k\bar{l}} (\Phi,\Phi^{\dagger}) 
\left(
\bar{\mathcal{D}}_{\dot{\alpha}} \Phi^{\dagger \bar{j}} e^{2gV}
 \mathcal{D}^{\alpha} \Phi^i
\right)
\left(
\bar{\mathcal{D}}^{\dot{\alpha}} \Phi^{\dagger \bar{l}} e^{2gV} 
\mathcal{D}_{\alpha} \Phi^k
\right)
\notag \\
& \ 
+ 
\left(
\int \! d^2 \theta \, W(\Phi) + \mathrm{h.c.}
\right)
+ \frac{1}{16 k g^2} 
\mathrm{Tr}
\left[
\int \! d^2 \theta \ W^{\alpha} W_{\alpha} + (\mathrm{h.c.})
\right]
- 2 \kappa g \int \! d^4 \theta \ \mathrm{Tr} V,
\label{eq:gauged_Lagrangian}
\end{align}
where we have introduced the gauge kinetic and Fayet-Iliopoulos (FI)
terms with the FI parameter $\kappa$.
The superpotential and the FI term are necessary to impose the supersymmetric constraints.
The bosonic part of the Lagrangian is given by \cite{Nitta:2015uba}
\begin{align}
\mathcal{L}_{\rm boson} =& \ 
- \frac{\partial^2 K}{\partial \bar{\varphi}_a^{\bar{j}} \partial \varphi^{ib}} 
D_m \bar{\varphi}_a^{\bar{j}} D^m \varphi^{ib}
- \frac{\partial^2 K}{\partial \bar{\varphi}_a^{\bar{j}} \partial \varphi^{ib}}
\bar{F}_a^{\bar{j}} F^{ib} + 
\frac{g}{2} D^{\hat{a}}
\left(
\bar{\varphi}_c^{\bar{j}} (T^{\hat{a}})^c {}_d \frac{\partial K}{\partial \bar{\varphi}_d^{\bar{j}}}
+
\frac{\partial K}{\partial \varphi^{ic}} (T^{\hat{a}})^c
 {}_d \varphi^{id} - 2 \kappa \delta^{\hat{a}} {}_0
\right)
\notag \\
& \ + \frac{1}{k} \mathrm{Tr} 
\left[
- \frac{1}{4} F_{mn} F^{mn} + \frac{1}{2} D^2
\right] 
+
\frac{\partial W}{\partial \varphi^{ia}} F^{ia} + \frac{\partial \bar{W}}{\partial \bar{\varphi}^{\bar{i}}_a} \bar{F}^{\bar{i}}_a 
\notag \\
& \ + \Lambda_{ik\bar{j}\bar{l}} (\varphi, \bar{\varphi}) 
\Bigg[
(D^m \bar{\varphi}_a^{\bar{j}} D^n \varphi^{ia}) (D_m \bar{\varphi}_b^{\bar{l}} D_n \varphi^{kb}) 
\notag \\
& \ \qquad \qquad \qquad \qquad 
- \frac{1}{2} (D_m \varphi^{ia} F^{kb} + F^{ia} D_m \varphi^{kb}) (D^m
 \bar{\varphi}_a^{\bar{j}} \bar{F}_b^{\bar{l}} + \bar{F}_a^{\bar{j}} D^m \bar{\varphi}_b^{\bar{l}}
 )
+ F^{ia} \bar{F}^{\bar{j}}_a F^{kb} \bar{F}^{\bar{l}}_b
\Bigg],
\label{eq:general_bosonic_Lagrangian}
\end{align}
where $F_{mn} = \partial_m A_n - \partial_m A_n + i [A_m, A_n]$ and 
$T^0$ is the $U(1)$ generator in $G'$.
A notable fact about the Lagrangian \eqref{eq:general_bosonic_Lagrangian} is that,
due to the derivative corrections, the equation of motion for the auxiliary field $F$ is not linear:
\begin{align}
&
- \frac{\partial^2 K}{\partial \bar{\varphi}^{\bar{j}'}_a \partial
 \varphi^{ib}} F^{ib}
+ \frac{\partial \bar{W}}{\partial \bar{\varphi}^{\bar{j}'}_a} 
\notag \\
&  
+ 
\Lambda_{ik\bar{j}\bar{l}} (\varphi, \bar{\varphi})
\Bigg[
- \frac{1}{2} 
\Big(
(D_m \varphi^{ib} F^{ka} + F^{ib} D_m \varphi^{ka})
\Big)
\Big(
D^m \varphi^{\bar{j}}_b \delta^{\bar{j}' \bar{l}} + \delta^{\bar{j}' \bar{j}} D^m \bar{\varphi}^{\bar{l}}_b
\Big)
\notag \\
& \qquad \qquad \qquad \qquad 
+ F^{ia} \delta^{\bar{j}'\bar{j}} F^{kb} \bar{F}^{\bar{l}}_b
+ F^{ib} \bar{F}^{\bar{j}}_b F^{ka} \delta^{\bar{j}' \bar{l}}
\Bigg] 
= 0.
\label{eq:auxiliary_eom}
\end{align}
This apparently allows several solutions for the auxiliary field.
The solutions provide several distinct on-shell Lagrangians.
For example, when $W = 0$, the inhomogeneous term in \eqref{eq:auxiliary_eom} vanishes and 
it is obvious that $F = 0$ is a solution.
We call this the canonical branch.
In addition, we have another solution $F \not= 0$ and we call the corresponding theory the non-canonical branch.
Both the branches exhibit remarkable features \cite{Nitta:2014pwa,SaYaYo, KoLeOv, FaKe}.
Although this shows an unusual situation in supersymmetric theories, 
we stress that Eq.~\eqref{eq:auxiliary_eom} is the algebraic equation,
not the 
kinematical one, which guarantees that $F$ still keeps the role of the auxiliary field.

In the following sections, we examine the nonlinear sigma model limit $g \to
\infty$ in the gauged linear sigma models.
In the limit, the vector multiplet 
carries non-propagating degrees of freedom and it will be integrated out.
We in particular focus on the Hermitian symmetric spaces of the type $G/H$ \cite{Higashijima:1999ki}.
This procedure makes us to write down a variety of supersymmetric nonlinear sigma
models with higher derivative terms.

\section{Nonlinear sigma models with D-term constraints}
\label{sect:NLSM}
In this section, we discuss supersymmetric nonlinear sigma models whose
target spaces are the complex projective space $\mathbb{C}P^{N-1}$ and
the Grassmann manifold $G_{M,N}$.
It is known that they are obtained in the sigma model limit $g \to
\infty$ of supersymmetric gauge theories with D-term constraints
\cite{DAdda:1978dle,Aoyama:1979zj,Higashijima:1999ki}.
The former is obtained from an Abelian gauge theory while the latter
comes from a non-Abelian gauge theory. 
In the following, we explicitly integrate out the gauge field both in
the canonical and the non-canonical branches in the sigma model limit $g
\to \infty$.

\subsection{$\mathbb{C}P^{N-1}$ model}

We first consider the nonlinear sigma model whose target
space is $\mathbb{C}P^{N-1} = SU(N) / [SU(N-1) \times U(1)]$.
We formulate this model as a gauged linear sigma model with 
the global symmetry $G = SU(N)$ and gauge symmetry $G'=U(1)$, 
where the latter is complexified  $G'_{\mathbb{C}} = U(1)_{\mathbb{C}}$.
We consider the flat K\"ahler potential $K= \delta_{i\bar{j}}
\Phi^i \Phi^{\dagger \bar{j}} = \Phi^{\dagger i} \Phi^i$ and $W = 0$.
The chiral superfields 
$\Phi^i, \Phi^{\dagger i}, (i = 1, \ldots, N)$ 
belong to the fundamental representation $\mathbf{N}$ 
of the global $SU(N)$ symmetry, and
their $U(1)$ charges are assigned as $(+1,-1)$.
We assume that the $(2,2)$ K\"ahler tensor is given by the flat metric as 
\begin{align}
 \Lambda_{ik \bar{j} \bar{l}} = \delta_{i\bar{j}} \delta_{k\bar{l}} \Lambda(\Phi,\Phi^\dagger),\end{align}
where $\Lambda (\Phi,\bar{\Phi})$ is a gauge invariant K\"ahler scalar
composed of the chiral superfields and their gauge covariant derivatives
$\mathcal{D}_{\alpha} \Phi$, $\bar{\mathcal{D}}_{\dot{\alpha}}
\Phi^{\dagger}$.\footnote{
The case of $\Lambda =$ const. was studied in Appendix B of Ref.~\cite{Eto:2012qda}.
}
Then, the Lagrangian becomes 
\cite{DAdda:1978dle}
\begin{align}
\mathcal{L} =& \ 
\int \! d^4 \theta \ \Phi^{\dagger i} e^{2V} \Phi^i + \frac{1}{16} 
\int \! d^4 \theta \ \Lambda (\Phi, \Phi^{\dagger}) \delta_{i \bar{j}}
 \delta_{k \bar{l}} \, e^{4V} \mathcal{D}^{\alpha} \Phi^i
 \mathcal{D}_{\alpha} \Phi^k \bar{\mathcal{D}}_{\dot{\alpha}}
 \Phi^{\dagger \bar{j}} \bar{\mathcal{D}}^{\dot{\alpha}} \Phi^{\dagger \bar{l}}
\notag \\
& 
+ \frac{1}{16 g^2}
\left[
\int \! d^2 \theta \ W^{\alpha} W_{\alpha} + (\mathrm{h.c.})
\right]
- 2  \kappa \int \! d^4 \theta \ V.
\label{eq:HDU1}
\end{align}
Here we have rescaled $V \to \frac{1}{g} V$.
We have introduced the FI term which provides the D-term constraints on
the component fields in $\Phi^i$.
Note that for the Abelian gauge group, we have $\mathcal{D}_{\alpha} \Phi = D_{\alpha} \Phi + (D_{\alpha} V)
\Phi$, $\bar{\mathcal{D}}_{\dot{\alpha}} \Phi^{\dagger} =
\bar{D}^{\dot{\alpha}} \Phi^{\dagger} + (\bar{D}^{\dot{\alpha}} V)
\Phi^{\dagger}$.
The bosonic part of the Lagrangian \eqref{eq:HDU1} is then given by
\begin{align}
\mathcal{L}_{\rm boson}  =& \ - D_m \varphi^i D^m \bar{\varphi}^{i} + 
F^i \bar{F}^i
+ D (\bar{\varphi}^i \varphi^i - \kappa) 
+ \frac{1}{2g} D^2 - \frac{1}{4g^2} F_{mn} F^{mn}
\notag \\
& \ 
+ 
\delta_{i\bar{j}} \delta_{k \bar{l}}
\Lambda (\varphi, \bar{\varphi})
\left(
\frac{}{}
(D_m \varphi^i D_n \bar{\varphi}^{\bar{j}}) (D^m \varphi^k D^n \bar{\varphi}^{\bar{l}})
- 2 (D_m \varphi^i D^m \bar{\varphi}^{\bar{j}}) (F^k \bar{F}^{\bar{l}})
+ F^i \bar{F}^{\bar{j}} F^k \bar{F}^{\bar{l}}
\right),
\label{eq:uv}
\end{align}
where $F_{mn} = \partial_m A_n - \partial_n A_m$ and the 
gauge covariant derivative is defined by $D_m \varphi^i = \partial_m \varphi^i + i A_m \varphi^i$.
In the following, we assume that the gauge invariant K\"ahler scalar $\Lambda (\varphi, \bar{\varphi})$
is given by a function of $X = D_m \varphi^i D^m \bar{\varphi}^i$ as a
typical example.
The most plausible reason for this assumption is that 
this quantity contains only the first order time derivative of fields.
This guarantees the absence of the Ostrogradski's ghost
\cite{Ostrogradski} in the theory.
This assumption is easily relaxed but allowing other 
gauge and Lorentz invariant quantities such as $X= \varphi^i
\bar{\varphi}^i, \, D_m \varphi^i D^n \bar{\varphi}^i
D^m \varphi^j D^n \bar{\varphi}^j$ and so on does not change the
following discussions.

In the sigma model limit $g \to \infty$, the gauge kinetic and $D^2$
terms vanish and the gauge field $A_m$ becomes an auxiliary field.
Before integrating out these fields, we first integrate out the auxiliary fields in the chiral superfields.
The equation of motion for the auxiliary field $\bar{F}^{\bar{i}}$ is 
\begin{align}
0 =& \ \frac{\partial \mathcal{L}}{\partial \bar{F}^{\bar{i}}} 
= F^i - 2 \Lambda (X)
\left(
(D_m \varphi^j D^m \bar{\varphi}^j) F^i - (F^j \bar{F}^j) F^i
\right).
\label{eq:auxiliary}
\end{align}
As we have noticed, there are two branches associated with the solutions
$F^i = 0$ and $F^i \not=0$. 
We examine each branch separately.

\paragraph{Canonical branch}
For the canonical branch corresponding to the solution $F^i = 0$,
the Lagrangian \eqref{eq:uv} becomes 
\begin{align}
\mathcal{L}_{\text{c}}  =& \ - D_m \varphi^i D^m \bar{\varphi}^{i} 
+ 
\Lambda 
(D_m \varphi^i D_n \bar{\varphi}^i) (D^m \varphi^j D^n
 \bar{\varphi}^j)
+ D
 (\bar{\varphi}^i \varphi^i - \kappa).
\label{eq:canonical_low_energy}
\end{align}
The equation of motion for the auxiliary field $D$ gives the following constraint for the scalar fields $\varphi^i$:
\begin{align}
\varphi^i \bar{\varphi}^i = \kappa.
\label{eq:constraint}
\end{align}
The equation of motion for the gauge field $A_m$ is 
\begin{align}
0 =  \frac{\partial \mathcal{L}}{\partial A_m} =& \  
-2 \kappa \left(
i \kappa^{-1} \varphi^i \partial^m \bar{\varphi}^i + A^m 
\right)
+ i \frac{\partial \Lambda}{\partial X} 
\left(
\varphi^i D^m \bar{\varphi}^i - \bar{\varphi}^i D^m \varphi^i
\right)
\notag \\
& \ 
+ 2 i \Lambda (X)
\Big[
(\varphi^i D_p \bar{\varphi}^i) (D_m \varphi^j D^p \bar{\varphi}^j)
- 
(\bar{\varphi}^i D_p \varphi^i) (D_m \bar{\varphi}^j D^p \varphi^j)
\Big].
\label{eq:gauge_eom}
\end{align}
Using the constraint \eqref{eq:constraint},
we find that the solution to this equation is given by 
\begin{align}
A_m =  i \kappa^{-1} \bar{\varphi}^i \partial_m \varphi^i.
\label{eq:gauge_sol}
\end{align}
Substituting the solution \eqref{eq:gauge_sol} into
\eqref{eq:canonical_low_energy} and using the constraint \eqref{eq:constraint},
we obtain the Lagrangian in the canonical branch:
\begin{align}
\mathcal{L}_{\text{c}}  =& \ - \tilde{D}_m \bar{\varphi}^i \tilde{D}^m \varphi^i 
+ 
\Lambda (\tilde{X})
(\tilde{D}_m \varphi^i \tilde{D}_n \bar{\varphi}^i) (\tilde{D}^m \varphi^j \tilde{D}^n \bar{\varphi}^j),
\label{eq:off-shell_Lagrangian}
\end{align}
where we have defined $\tilde{D} \varphi^i = \partial_m \varphi^i -
\kappa^{-1} (\bar{\varphi}^j \partial_m \varphi^j) \varphi^i$ and
$\tilde{X} = \tilde{D}_m \varphi^i \tilde{D}^m \bar{\varphi}^i$.

It is convenient to solve the constraint \eqref{eq:constraint} explicitly by using the 
parametrization 
\begin{align}
\varphi^i = W^i \frac{\sqrt{\kappa}}{\sqrt{W^{\dagger} \cdot W}}.
\end{align}
By using the $U(1)_{\mathbb{C}}$ gauge symmetry, we can fix $W^1 = 1$. 
Then, we have the conventional parametrization of the $\mathbb{C}P^{N-1}$ model:
\begin{align}
\varphi^i = \frac{\sqrt{\kappa}}{\sqrt{1 + |\vec{u}|^2}}
\left(
\begin{array}{c}
1 \\
u^{s}
\end{array}
\right),
\qquad 
\bar{\varphi}^{\bar{i}} = \frac{\sqrt{\kappa}}{\sqrt{1 + |\vec{u}|^2}}
\left(
\frac{}{}
1,\  \bar{u}^{\bar{s}}
\right), \qquad 
(s,\bar{s} = 2, \ldots, N),
\label{eq:u}
\end{align}
where $u^{s}, \bar{u}^{\bar{s}}$ are inhomogeneous coordinates of the ${\mathbb C}P^{N-1}$ manifold.
By using this form, the Lagrangian in the canonical branch can be rewritten as 
\begin{align}
\mathcal{L}_{\text{c}} =& \ - \frac{\kappa}{(1 + |\vec{u}|^2)^2}
\left[
(1 + |\vec{u}|^2) (\partial_m \vec{u}^{*} \cdot \partial^m \vec{u})
-
 (\vec{u} \cdot \partial_m \vec{u}^{*})
(\vec{u}^{*} \cdot \partial^m \vec{u})
\right] \notag \\
& \ + 
\frac{\kappa^2}{(1 + |\vec{u}|^2)^4} \Lambda (u,\bar{u})
\left[ \frac{}{}
(1 + |\vec{u}|^2)^2
(\partial_m \vec{u}^{*} \cdot \partial_n \vec{u}) (\partial^m
 \vec{u}^{*} \cdot \partial^n \vec{u})
\right.
\notag \\
& \ \qquad \quad 
\left.
\frac{}{}
- 2 (1 + |\vec{u}|^2) 
(\partial_m \vec{u}^{*} \cdot \partial_n
 \vec{u} ) (\vec{u} \cdot \partial^m \vec{u}^{*})
 (\vec{u}^{*} \cdot \partial^n \vec{u})
+ (\vec{u} \cdot \partial_m \vec{u}^{*})^2
(\vec{u}^{*} \cdot \partial_n \vec{u})^2
\right].
\label{eq:canonical_u_lagrangian}
\end{align}
By using the Fubini-Study metric
\begin{align}
g_{s\bar{t}} = \kappa \frac{(1 + |\vec{u}|^2) \delta_{s\bar{t}} - u^{s} \bar{u}^{\bar{t}}}{(1
 + |\vec{u}|^2)^2},
\label{eq:Fubini-Study}
\end{align}
 for the $\mathbb{C}P^{N-1}$ manifold, 
the Lagrangian \eqref{eq:canonical_u_lagrangian} can be simply written as
\begin{align}
\mathcal{L}_{\text{c}} = - g_{s\bar{t}} \partial_m u^{s} \partial^m
 \bar{u}^{\bar{t}} 
+ g_{s\bar{t}} g_{q\bar{r}} 
\Lambda (u,\bar{u}, \partial_m u, \partial_m \bar{u}) 
(\partial_m u^{s} \partial_n \bar{u}^{\bar{t}})
 (\partial^m u^{q} \partial^n \bar{u}^{\bar{r}}).
 \label{eq:CPN}
\end{align}
The first term is the ordinary kinetic term of the $\mathbb{C}P^{N-1}$ nonlinear sigma model
and the second is the derivative corrections determined by the arbitrary
scalar function $\Lambda (u,\bar{u}, \partial_m u, \partial_m
\bar{u})$.

Before going to the discussion on the the non-canonical branch,
two comments in this Lagrangian are in order.
This Lagrangian with $\Lambda=$const. 
was obtained in Ref.~\cite{Eto:2012qda}
as 
the low-energy effective theory of 
a BPS non-Abelian vortex in ${\cal N}=2$ supersymmetric $U(N)$ gauge theory 
\cite{Hanany:2003hp};
A non-Abelian vortex in this case allows 
the supersymmetric ${\mathbb C}P^{N-1}$ model on the vortex worldsheet,
on which 
1/2 BPS condition preserves four supercharges among 
 eight supercharges 
that the ${\cal N}=2$ supersymmetric theory in the bulk has.
The bosonic part of the four-derivative correction term is 
 precisely in this form \cite{Eto:2012qda}.

Without supersymmetry, the ${\mathbb C}P^{N-1}$ Skyrme-Faddeev model 
was proposed in 
Refs.~\cite{Ferreira:2010jb,Ferreira:2011jy,Ferreira:2011ja} 
as  the ${\mathbb C}P^{N-1}$ model with four derivative terms.
In this case, there are three kinds of  four derivative terms, 
and the one in Eq.~(\ref{eq:CPN}) (with $\Lambda=$const.)
corresponds to a particular choice
among them. 
Thus, we call the Lagrangian in Eq.~(\ref{eq:CPN}) 
(the bosonic part of) the supersymmetric ${\mathbb C}P^{N-1}$ Skyrme-Faddeev model.

\paragraph{Non-canonical branch}
We next study the non-canonical branch corresponding to solutions $F_i
\not= 0$.
Again, the equation of motion for $\bar{F}$ is 
\begin{align}
0 =& \ \frac{\partial \mathcal{L}}{\partial \bar{F}^{\bar{i}}} 
=  F^i - 2 \Lambda (X)
\Big(
( D_m \varphi^j D^m \bar{\varphi}^j) F^i - (F^j \bar{F}^j) F^i
\Big).
\end{align}
The solution $F^i \not= 0$ is found to be
\begin{align}
F^i \bar{F}^i = 
- \frac{1}{2 \Lambda} + D_m \varphi^i D^m \bar{\varphi}^i.
\end{align}
Substituting this solution into the Lagrangian \eqref{eq:uv}, it becomes
\begin{align}
\mathcal{L}_{\text{nc}}  = 
\Lambda (X)
\left[
(D_m \bar{\varphi}^i D_n \varphi^i) (D^m \bar{\varphi}^j D^n \varphi^j)
-
(D_m \varphi^j D^m \bar{\varphi}^j)^2
\right]
- \frac{1}{4 \Lambda (X)}
+ D (\bar{\varphi}^i \varphi^i - \kappa).
\label{eq:non-canonical_low_energy}
\end{align}
where we have assumed the sigma model limit $g \to \infty$.
The D-term condition leads to the constraint \eqref{eq:constraint} while 
the equation of motion for the gauge field is 
\begin{align}
0 =& \ 
i \frac{\partial \Lambda}{\partial X} 
\left(
\varphi^k D^m \bar{\varphi}^k - \bar{\varphi}^k D^m \varphi^k
\right) 
\left[
(D_m \bar{\varphi}^i D_n \varphi^i) (D^m \bar{\varphi}^j D^n \varphi^j)
-
(D_m \varphi^j D^m \bar{\varphi}^j)^2
\right]
\notag \\
& \ 
+
2 i \Lambda (X)
\Big[\frac{}{}
- \bar{\varphi}^j D^q \varphi^j (D^m \bar{\varphi}^i D_q \varphi^i)
+ \varphi^j D^p \bar{\varphi}^j (D_p \bar{\varphi}^i D^m \varphi^i)
\notag \\
& \ 
\qquad \qquad \qquad \qquad \qquad 
- (D_p \varphi^i D^p \bar{\varphi}^i) (\varphi^j D^m \bar{\varphi}^j -
 \bar{\varphi}^j D^m \varphi^j)
\Big]
\notag \\
& \ 
+ 4 i \Lambda^{-2} (X) \frac{\partial \Lambda}{\partial X} 
\left(
\varphi^i D^m \bar{\varphi}^i - \bar{\varphi}^i D^m \varphi^i
\right).
\label{eq:cpn_nc}
\end{align}
By using the constraint \eqref{eq:constraint}, 
we find that the equation \eqref{eq:cpn_nc} is again solved by 
\begin{align}
A_m =  i \kappa^{-1} \bar{\varphi}^i \partial_m \varphi^i.
\label{eq:cpn_nc_gauge_sol}
\end{align}
Substituting this solution into the Lagrangian \eqref{eq:non-canonical_low_energy}, we find 
the new sigma model given by
\begin{align}
\mathcal{L}_{\text{nc}}  = 
\Lambda (\tilde{X})
\left[
(\tilde{D}_m \bar{\varphi}^i \tilde{D}_n \varphi^i) (\tilde{D}^m \bar{\varphi}^j \tilde{D}^n \varphi^j)
-
(\tilde{D}_m \varphi^j \tilde{D}^m \bar{\varphi}^j)^2
\right]
- \frac{1}{4 \Lambda(\tilde{X})}.
\label{eq:cpn_ncb_lagrangian}
\end{align}
The expression \eqref{eq:u} enable us to rewrite the Lagrangian
\eqref{eq:cpn_ncb_lagrangian} as 
\begin{align}
\mathcal{L}_{\text{nc}} =& \ \frac{\kappa^2}{(1 + |\vec{u}|^2)^4}
 \Lambda (u,\bar{u}, \partial_m u, \partial_m \bar{u}) 
\left[
\frac{}{}
(1 + |\vec{u}|^2)^2 
\left\{
(\partial_m \vec{u} \cdot \partial_n \vec{u}^*) (\partial^m \vec{u}
 \cdot \partial^n \vec{u}^*) - (\partial_m \vec{u}^{*} \cdot \partial^m \vec{u})^2
\right\}
\right.
\notag \\
& \qquad 
- 2 (1 + |\vec{u}|^2) 
\left\{
(\partial_m \vec{u}^{*} \cdot \partial_n \vec{u})
(\vec{u} \cdot \partial^m \vec{u}^{*})
(\vec{u}^{*} \cdot \partial^n \vec{u})
-
(\partial_m \vec{u}^{*} \cdot \partial^m \vec{u})
(\vec{u} \cdot \partial_n \vec{u}^{*})
(\vec{u}^{*} \cdot \partial^n \vec{u})
\right\}
\notag \\
& \qquad 
\left.
+ (\vec{u} \cdot \partial_m \vec{u}^{*})^2 (\vec{u}^{*} \cdot \partial_n
 \vec{u})^2 - 
\left\{
(\vec{u} \cdot \partial_m \vec{u}^{*}) (\vec{u}^{*} \cdot \partial^m \vec{u})
\right\}^2
\right] - \frac{1}{4}\Lambda^{-1} (u,\bar{u}, \partial_m u, \partial_m \bar{u}).
\end{align}
By using the Fubini-Study metric \eqref{eq:Fubini-Study}, 
this can be simply expressed as 
\begin{align}\label{eq:CPN-NC}
\mathcal{L}_{\text{nc}} =& \ \Lambda (u,\bar{u}, \partial_m u, \partial_m \bar{u}) 
\left[
\frac{}{}
(
g_{s \bar{t} } \partial_m u^{s} \partial_n \bar{u}^{\bar{t}}
)
(
g_{q \bar{r}} \partial^m u^{q} \partial^n \bar{u}^{\bar{q}}
)
-
(g_{s \bar{t}} \partial_m u^{s} \partial^m \bar{u}^{\bar{t}})^2
\right] \\\nonumber
& - \frac{1}{4} \Lambda^{-1} (u,\bar{u}, \partial_m u, \partial_m \bar{u}).
\end{align}
This four derivative term does not contain fourth order time
derivatives, unlike the one in Eq.~(\ref{eq:CPN}) in the canonical branch.
It is obvious that the target space of the scalar
fields is the $\mathbb{C}P^{N-1}$ space but the canonical kinetic 
term $g_{s \bar{t}}\del_m u^{s} \del^m u^{\bar{t}}$ is absent.

In particular for the $N=2$ case of $\mathbb{C}P^1$ , the Lagrangian reduces to
\begin{align}
\mathcal{L}_{\text{nc}} =& \ 
- \kappa^2 \Lambda \frac{(\partial_m u \partial^m \bar{u})^2 - |\partial_m u \partial^m
 u|^2}{(1 + |u|^2)^4} - \frac{1}{4 \Lambda}
\end{align}
which is known as the supersymmetric baby Skyrme model 
\cite{Adam:2011hj,AdQuSaGuWe,Nitta:2014pwa,Bolognesi:2014ova,Queiruga:2016jqu,Gudnason:2020tps}.
One notices that the first term 
(with $\Lambda=1$)
is nothing but the Skyrme-Faddeev fourth derivative term 
(without the fourth order time derivatives)
while the second term provides higher derivative corrections determined
by $\Lambda (u,\bar{u}, \partial_m u, \partial_m \bar{u})$.
The Lagrangian in Eq.~(\ref{eq:CPN-NC}) may be called 
 the supersymmetric ${\mathbb C}P^{N-1}$ baby Skyrme model.

\subsection{$G_{M,N}$ model}
We next consider the nonlinear sigma model 
whose target space is the Grassmann manifold 
$G_{M,N} = U(M)/[U(M-N) \times U(N)]$.
We consider a gauged linear sigma model 
with the global symmetry $G = SU(M)$ 
and gauge symmetry $U(N)$,
of which 
the chiral superfield $\Phi^i_a$ 
($i = 1, \ldots M$; $a = 1, \ldots N$) belong to the $(\mathbf{M},\bar{\mathbf{N}})$
representation. 
The global symmetry $G = SU(M)$ 
and gauge symmetry $U(N)_{\mathbb C} = GL(N,{\mathbb C})$ act on it as
\begin{align}
\Phi \to 
g \Phi e^{i \Theta'},
\qquad 
e^{2V} \to e^{- i \Theta'} e^{2V} e^{i \Theta^{\prime \dagger}},
\end{align}
where $g \in SU(M)$,  
$\Theta'(x,\theta,\bar \theta)$ is a chiral superfield of 
a $U(N)_{\mathbb C}$ gauge parameter, 
and we have introduced 
the associated $U(N)$ vector superfield
$V(x,\theta,\bar\theta)$.
The $G \times G'$ invariant Lagrangian is therefore 
\cite{Aoyama:1979zj,Higashijima:1999ki}
\begin{align}
\mathcal{L} =& \ \int \! d^4 \theta \, 
\mathrm{Tr} 
\left[
\Phi e^{2V} \Phi^{\dagger}
\right]
- \frac{1}{16} \int \! d^4 \theta \, \Lambda (\Phi,\Phi^{\dagger})
\mathrm{Tr}
\left[
\mathcal{D}^{\alpha} \Phi e^{2V} \bar{\mathcal{D}}_{\dot{\alpha}} \Phi^{\dagger}
\right]
\mathrm{Tr}
\left[
\mathcal{D}_{\alpha} \Phi e^{2V} \bar{\mathcal{D}}^{\dot{\alpha}} \Phi^{\dagger}
\right]
\notag \\
& \ 
+ \frac{1}{16g^2} 
\left(
\int \! d^2 \theta \, \mathrm{Tr} [ W^{\alpha} W_{\alpha} ] + (\mathrm{h.c.})
\right) 
- 2 \kappa \int \! d^4 \theta \, \mathrm{Tr} V.
\end{align}
The derivative corrections are given in the double trace form and 
we have included the FI term
which provides the D-term conditions on the chiral multiplets.

In the nonlinear sigma model limit $g \to \infty$, the bosonic part of the Lagrangian becomes
\begin{align}
\mathcal{L}_{\rm boson} =& \ - \mathrm{Tr} [D_m \varphi D^m
 \bar{\varphi}] + \mathrm{Tr} [\bar{\varphi} D \varphi - \kappa D] + \mathrm{Tr} [F\bar{F}]
\notag \\
& \ + \Lambda (\varphi, \bar{\varphi})
\Big\{
\mathrm{Tr} [D_m \varphi D_n \bar{\varphi}] \mathrm{Tr} [D^m \varphi D^n
 \bar{\varphi}] 
- \mathrm{Tr} [D_m \varphi D^m \bar{\varphi}] \mathrm{Tr} [F \bar{F}]
\notag \\
& \ \qquad \qquad \qquad 
- \mathrm{Tr} [F D_m \bar{\varphi}] \mathrm{Tr} [D^m \varphi
 \bar{F}]
+ \mathrm{Tr} [F \bar{F}] \mathrm{Tr} [F \bar{F}]
\Big\}.
\end{align}
The gauge covariant derivative is $D_m \varphi = \partial_m \varphi + i
\varphi A_m$.
In the following, we assume $\Lambda$ is given by the gauge invariant
quantity $X = \mathrm{Tr} [D_m \varphi D^m \bar{\varphi}]$.
As clarified before, this condition is easily relaxed allowing for any
other gauge invariant quantities.
The equation of motion for the auxiliary field $D$ gives the constraint:
\begin{align}
\bar{\varphi} \varphi = \kappa \mathbf{1}_N.
\label{eq:D-constraint2}
\end{align}
On the other hand, the equation of motion for the auxiliary field
$\bar{F}$ is given by 
\begin{align}
F + \Lambda (X) 
\Big\{
- \mathrm{Tr} [D_m \varphi D^m \bar{\varphi}] F - \mathrm{Tr} [F D_m
 \bar{\varphi}] D^m \varphi + 2 \mathrm{Tr} [F \bar{F}] F
\Big\} = 0.
\label{eq:eqF2}
\end{align}

\paragraph{Canonical branch}
It is obvious that $F = 0$ is a solution.
In this case, the Lagrangian becomes
\begin{align}
\mathcal{L}_{\mathrm{c}} =& \ - \mathrm{Tr} [D_m \varphi D^m
 \bar{\varphi}] + \Lambda (X) 
\mathrm{Tr} [D_m \varphi D_n \bar{\varphi}] \mathrm{Tr} [D^m \varphi D^n
 \bar{\varphi}],
\end{align}
where the scalar fields satisfy the constraint \eqref{eq:D-constraint2}.
The equation of motion for $A_m$ then becomes,
\begin{align}
&
\bar{\varphi} \partial_m \varphi + i \kappa A_m
+ i \frac{\partial \Lambda}{\partial X} 
\left(
\varphi D^m \bar{\varphi} - D^m \varphi \bar{\varphi}
\right)
\notag \\
& \qquad 
- \Lambda (X) 
\mathrm{Tr} 
\Big[
D_m \varphi D_n \bar{\varphi}
+
D_n \varphi D_m \bar{\varphi}
\Big]
\left(
\bar{\varphi} \partial^n \varphi + i \kappa A^n
\right) = 0,
\end{align}
where we have used the constraint \eqref{eq:D-constraint2}.
We again find that the exact solution to this equation is given by
\begin{align}
A_m = i \kappa^{-1} \bar{\varphi} \partial_m \varphi.
\end{align}
Plugging this solution back into the Lagrangian, we obtain the
nonlinear sigma model supplemented by the derivative corrections:
\begin{align}
\mathcal{L}_{\mathrm{c}} =& \ - \mathrm{Tr} [\tilde{D}_m \varphi
 \tilde{D}^m \bar{\varphi}] + \Lambda (\tilde{X})
 \mathrm{Tr} [\tilde{D}_m \varphi \tilde{D}_n \bar{\varphi}] \mathrm{Tr}
 [\tilde{D}^m \varphi \tilde{D}^n \bar{\varphi}].
\label{eq:Grassmann_Lagrangian}
\end{align}
Here $\tilde{D}_m \varphi = \partial_m \varphi - \kappa^{-1} \varphi
(\bar{\varphi} \partial_m \varphi)$ and $\tilde{X} = \mathrm{Tr}
[\tilde{D}_m \varphi \tilde{D}^m \bar{\varphi}]$.
The constraint \eqref{eq:D-constraint2} is solved by the
parametrization:
\begin{align}
\varphi = W \frac{\sqrt{\kappa}}{\sqrt{W^{\dagger} W}},
\label{eq:grassmann_parametrization}
\end{align}
where $W$ is an $M \times N$ matrix.
Substituting this into the Lagrangian, the first term in
\eqref{eq:Grassmann_Lagrangian} becomes the kinetic term of the sigma model
whose target space is the Grassmann manifold $G_{M,N}$ while the second
term gives derivative corrections determined by the arbitrary gauge
invariant function $\Lambda (W,W^{\dagger}, \del_m W, \del_m W^{\dagger})$.

\paragraph{Non-canonical branch}
We next examine the non-canonical branch.
It is obvious that there are non-zero solutions $F \not= 0$ to the equation
\eqref{eq:eqF2}.
Compared with the case of the Abelian gauge symmetry, it is not
straightforward to write down explicit
solutions for $F$. However, the gauge invariant solution $\mathrm{Tr} [F \bar{F}]$ should be given by the gauge and
the Lorentz invariant quantities $Y$ such as 
\begin{align}
Y = \mathrm{Tr} [D_m \varphi D^m \bar{\varphi}], \quad 
\mathrm{Tr} [D_m \varphi D_n \bar{\varphi} D^m \varphi D^n
 \bar{\varphi}], \quad 
\mathrm{Tr} [D_m \varphi D_n \bar{\varphi} D^n \varphi D^m
 \bar{\varphi}],
\label{eq:gauge_inv_quantities}
\end{align}
and so on. Using the equation of motion for $F$, the non-canonical Lagrangian
becomes 
\begin{align}
\mathcal{L}_{\mathrm{nc}}  = 
- \mathrm{Tr} [D_m \varphi D^m \bar{\varphi}] 
+ \Lambda (X)
\Big\{
\mathrm{Tr} [D_m \varphi D_n \bar{\varphi}] \mathrm{Tr} [D^m \varphi D^n
 \bar{\varphi}]
- \mathrm{Tr} [F\bar{F}(Y)]^2
\Big\}.
\label{eq:dc_grassmann}
\end{align}
Here $\mathrm{Tr} [F \bar{F} (Y)]$ is a solution $F \not= 0$
given by $Y$.
The variation of the Lagrangian \eqref{eq:dc_grassmann} by $A_m$ is therefore
\begin{align}
\delta \mathcal{L}_{\mathrm{nc}} =& \ 
i \mathrm{Tr} 
\left[
(\bar{\varphi} \partial^m \varphi + i \kappa A^m) \delta A_m
\right]
+
i \frac{\partial \Lambda}{\partial X} 
\mathrm{Tr}
\Big[
\left(
\varphi D^m \bar{\varphi} - D^m \varphi \bar{\varphi}
\right) \delta A_m
\Big]
\notag \\
& \ 
- i \Lambda (X) 
\mathrm{Tr} 
\Big[
D^m \varphi D^n \bar{\varphi} + D^n \varphi D^m \bar{\varphi}
\Big]
\mathrm{Tr} 
\Big[
(\bar{\varphi} \partial_n \varphi + i \kappa A_n) \delta A_m
\Big]
\notag \\
& \ 
 - 2 \Lambda (\varphi, \bar{\varphi}) \mathrm{Tr} [F \bar{F}]
\mathrm{Tr} 
\left[
(F\bar{F})' \delta Y 
\right].
\label{eq:variation_ncb_grassmann}
\end{align}
Here $(F\bar{F})'$ stands for the differentiation with respect to $Y$.
In the last term in \eqref{eq:variation_ncb_grassmann}, 
one finds that the variation of the gauge invariant quantity $Y$ is
proportional to $\bar{\varphi} \partial^m \varphi + i \kappa A^m$.
Indeed, for $Y = \mathrm{Tr} [D_m \varphi D^m \bar{\varphi}]$, we
have
\begin{align}
\delta Y = 2i \mathrm{Tr} [(\bar{\varphi} \partial^m \varphi + i \kappa
 A^m) \delta A_m].
\end{align}
Therefore even in the non-canonical branch, we find that the solution to
the equation of motion for $A_m$ is given by $A_m = i \kappa^{-1}
\varphi \partial_m \bar{\varphi}$.
For the other gauge invariant combinations $Y$ such as \eqref{eq:gauge_inv_quantities},
the result is the same.
The constraint \eqref{eq:D-constraint2} is solved by the parametrization
\eqref{eq:grassmann_parametrization} and the Lagrangian becomes again
the Grassmanian sigma model with the derivative corrections.

We comment on the symmetric nature of the Grassmanian manifold
$G_{M,N} = G_{M,M-N}$.
We have constructed the Grassmanian sigma models by gauging the symmetry
$G' = U(N)$ but there is an alternative way to define the same target
space by gauging $G' = U(M-N)$.
It is obvious that the latter construction is essentially the same with
the former one except the $N=1$ case.
In this case, one notices that $G_{N,1} = G_{N,N-1} = \mathbb{C}P^{N-1}$.
This equality implies that we can construct supersymmetric sigma model
whose target space is $\mathbb{C}P^{N-1}$, either by gauging the
non-Abelian $U(N-1)$ symmetry or the Abelian $U(1)$ symmetry.
The former is nothing but the construction discussed in this section
while the latter is the formulation discussed in the
previous section.
We find apparently different Lagrangians \eqref{eq:dc_grassmann} and
\eqref{eq:cpn_ncb_lagrangian} in the derivative corrections.
This means that the equivalence of the target spaces in different
formalism are not taken over to the derivative corrections in the
superfield language.
In the next section, 
we will see a similar situation in the canonical branch of the nonlinear
sigma models defined by the F-term constraints. 
We will, among other things, encounter qualitatively different structures of the
target spaces in the non-canonical branch.

\section{Nonlinear sigma models with D- and F-term constraints} \label{sect:F-term}
In this section, we discuss the nonlinear sigma models defined 
by F-term constraints in addition to the D-term constraints
\cite{Higashijima:1999ki,Higashijima:2000rp}.
Such Hermitian symmetric target spaces include the quadric surface $Q^{N-2} = SO(N)/[SO(N-2)
\times U(1)]$, the quotient spaces of the types $SO(2N)/U(N)$,
$Sp(2N)/U(N)$, $E_6/[SO(10) \times U(1)]$ and $E_7/[E_6 \times U(1)]$.
We begin with the target space $Q^{N-2} = SO(N)/[SO(N-2)
\times U(1)]$ as a typical example.

\subsection{Quadric surface $Q^{N-2}$}
  We consider a gauged linear sigma model with 
the global symmetry $G = SO(N)$ and the gauge symmetry $U(1)$.
The chiral superfield $\Phi^i$ ($i = 1, \ldots, N$) belongs to the fundamental representation
$\mathbf{N}$ of $SO(N)$ and has the $U(1)$ charge $+1$.
The Lagrangian in the nonlinear sigma model limit is 
\begin{align}
\mathcal{L} =& \  \int \! d^4 \theta \, 
\left(
\Phi^i e^{2V} \Phi^{\dagger i} - 2 \kappa V
\right)
+
\left(
 \int \! d^2 \theta \, 
\Phi_0 \vec{\Phi}^t J \vec{\Phi}
+ \mathrm{h.c.}
\right)
\notag \\
& \ 
+ \frac{1}{16} \int \! d^4 \theta \, \Lambda (\Phi, \Phi^{\dagger}) 
e^{4V} \mathcal{D}^{\alpha} \Phi^i \mathcal{D}_{\alpha} \Phi^j 
\bar{\mathcal{D}}_{\dot{\alpha}} \Phi^{\dagger i}
 \bar{\mathcal{D}}^{\dot{\alpha}} \Phi^{\dagger j},
\label{eq:QN_superfield_Lagrangian}
\end{align}
where we have introduced the gauge invariant superpotential $W = \Phi_0
\vec{\Phi}^t J \vec{\Phi}$.
This will provide the F-term constraints on the chiral multiplet.
Here $\Phi_0$ is the $SO(N)$ singlet and whose $U(1)$ charge is assigned
to $-2$, $V$ is the $U(1)$ vector multiplet and the matrix $J$ is given by 
\begin{align}
J = 
\left(
\begin{array}{ccc}
0 & 0 & 1 \\
0 & {\bf 1}_{N-2} & 0 \\
1 & 0 & 0
\end{array}
\right).
\end{align}
The gauge transformations are given by  
\begin{align}
\Phi \to e^{- i \Phi'} \Phi, \quad \Phi_0 \to e^{2 i \Theta'} \Phi_0,
 \quad e^{2V} \to e^{- i \Theta^{\prime \dagger}} e^{2V} e^{i \Theta'}.
\end{align}
Then the bosonic part of the Lagrangian is evaluated as 
\begin{align}
\mathcal{L}_{\mathrm{boson}} =& \ - D_m \varphi^i D^m \bar{\varphi}^i + F^i
 \bar{F}^i 
+ F_0 (\vec{\varphi}^t J \vec{\varphi}) + \bar{F}_0
 (\vec{\bar{\varphi}}^t J \vec{\bar{\varphi}})
+ D(\bar{\varphi}^i \varphi^i - \kappa)
\notag \\
& \ 
+ \Lambda (X) 
\Big\{
(D_m \varphi^i D_n \bar{\varphi}^i) (D^m \varphi^j D^n \bar{\varphi}^j)
- 2 (D_m \varphi^i D^m \bar{\varphi}^i) F^j \bar{F}^j + (F^i F^i)^2
\Big\}
\notag \\
& \ 
+ 2 \varphi_0 
\left\{
\varphi^1 F^N + \varphi^N F^1 + \sum_{\hat{i}=2}^{N-1} \varphi^{\hat{i}} F^{\hat{i}}
\right\}
+ 2 \bar{\varphi}_0
\left\{
\bar{\varphi}^1 \bar{F}^N + \bar{\varphi}^N \bar{F}^1 
+ \sum_{\hat{i}=2}^{N-1} \bar{\varphi}^{\hat{i}} \bar{F}^{\hat{i}}
\right\},
\end{align}
where we have again assumed that $\Lambda$ is a function of the gauge
invariant quantity $X = D_m \varphi^i D^m \bar{\varphi}^i$.
The equation of motion for $D$ gives the constraint
\begin{align}
\bar{\varphi}^i \varphi^i = \kappa.
\label{eq:Qc0}
\end{align}
Since $\Phi_0$ does not propagate, it is also integrated out.
The F-term constraints are given by the equation of motions for $F_0, \varphi_0$:
\begin{align}
&
\delta F_0 : \vec{\varphi}^t J \vec{\varphi} = 0, 
\label{eq:Qc1}
\\
&
\delta \varphi_0 : \varphi^1 F^{N} + \varphi^N F^1 + \sum_{\hat{i}=2}^{N-1}
 \varphi^{\hat{i}} F^{\hat{i}} = 0.
\label{eq:Qc2}
\end{align}
These give the constraints for the fields $\varphi^i, \bar{\varphi}^i$.
The equations of motion for $\bar{F}^i$ are 
\begin{align}
&
\delta \bar{F}^1 : 
F^1 - 2 \Lambda (X) 
\Big\{
(D_m \varphi^j D^m \bar{\varphi}^j) F^1 - (F^j \bar{F}^j) F^1
\Big\} + 2 \bar{\varphi}_0 \bar{\varphi}^N = 0,
\notag \\
&
\delta \bar{F}^{\hat{i}} :
F^{\hat{i}} - 2 \Lambda (X) 
\Big\{
(D_m \varphi^j D^m \bar{\varphi}^j) F^{\hat{i}} - (F^j \bar{F}^j) F^{\hat{i}}
\Big\}
+ 2 \bar{\varphi}_0 \bar{\varphi}^{\hat{i}} = 0, \quad  (\hat{i}=2,\ldots,N-1),
\notag \\
&
\delta \bar{F}^{N} : 
F^N - 2 \Lambda (X) 
\Big\{
(D_m \varphi^j D^m \bar{\varphi}^j) F^N - (F^j \bar{F}^j) F^N 
\Big\}
+ 2 \bar{\varphi}_0 \bar{\varphi}^1 = 0.
\label{eq:eqF3}
\end{align}
The equations \eqref{eq:Qc0}, \eqref{eq:Qc1}, \eqref{eq:Qc2} and \eqref{eq:eqF3} 
should be solved simultaneously.
In the following, we solve these equations in the canonical and the non-canonical branches separately.

\paragraph{Canonical branch}
One finds that a solution to the equations \eqref{eq:eqF3} is given by 
\begin{align}
\varphi_0 = 0, \quad F^i = 0.
\end{align}
This corresponds to the canonical branch.
We note that in this branch, the constraint \eqref{eq:Qc2} becomes trivial.
Then the Lagrangian becomes
\begin{align}
\mathcal{L}_{\mathrm{c}} =& \ 
- D_m \varphi^i D^m \bar{\varphi}^i 
+ 
\Lambda (X) (D_m \varphi^i D_n \bar{\varphi}^i)
 (D^m \varphi^j D^n \bar{\varphi}^j).
\label{eq:QN_canonical}
\end{align}
The equation of motion for $A_m$ is therefore
\begin{align}
&
- \kappa (i \kappa^{-1} \varphi^i \partial_m \bar{\varphi}^i + A_m)
+ i \frac{\partial \Lambda}{\partial X} (\varphi^i D^m \bar{\varphi}^i -
 \bar{\varphi}^i D^m \varphi^i)
\notag \\
& 
+ i \Lambda (\varphi, \bar{\varphi})
\Bigg\{
(\varphi^i D_n \bar{\varphi}^i) (D_m \varphi^j D^n \bar{\varphi}^j)
- (\bar{\varphi}^i D^n \varphi^i) (D_m \bar{\varphi}^j D^n \varphi^j)
\Bigg\} = 0,
\end{align}
where we have used the constraint \eqref{eq:Qc0}.
We find that the solution is given by
\begin{align}
A_m = i \kappa^{-1} \bar{\varphi}^i \partial_m \varphi^i.
\end{align}
Substituting this solution to the Lagrangian, 
the covariant derivative $D_m \varphi^i$ in the Lagrangian \eqref{eq:QN_canonical} 
is replaced by $\tilde{D}_m \varphi^i$.
Now we solve the remaining constraints \eqref{eq:Qc0} and
\eqref{eq:Qc1}.
The constraint \eqref{eq:Qc0} is solved by the following parametrization:
\begin{align}
\varphi^i = W^i \frac{\sqrt{\kappa}}{\sqrt{W^{\dagger} \cdot W}}.
\label{eq:Qphi1}
\end{align}
Since we have
\begin{align}
\vec{\varphi}^t J \vec{\varphi} = \frac{\kappa}{W^{\dagger} \cdot W} 
(2 W^1 W^N + (W^{\hat{i}})^2) = 0,  \qquad (\hat{i} = 2, \ldots, N-1),
\end{align}
the last constraint \eqref{eq:Qc1} is solved by
\begin{align}
W^i = 
\left(
\begin{array}{c}
1  \\
u^{\hat{i}} \\
 - \frac{1}{2} (u^{\hat{i}})^2 
\end{array}
\right),
\label{eq:Qphi2}
\end{align}
where we have fixed $W^1 = 1$ by the $U(1)_{\mathbb{C}}$ gauge transformation.
Plugging this back into the Lagrangian, we obtain the $Q^{N-2}$ sigma
model whose derivative corrections are completely determined by $\Lambda (W,
W^{\dagger}, \del_m W, \del_m W^{\dagger})$.

\paragraph{Non-canonical branch}
In the non-canonical branch, we find that a solution to the equations \eqref{eq:eqF3} is
given by
\begin{align}
\varphi_0 = 0, \quad F^i \bar{F}^i = - \frac{1}{2 \Lambda (X)} + D_m
 \varphi^i D^m \bar{\varphi}^i.
\end{align}
Then, the Lagrangian becomes 
\begin{align}
\mathcal{L}_{\mathrm{nc}} = 
\Lambda (X)
\left[
(D_m \bar{\varphi}^i D_n \varphi^i) (D^m \bar{\varphi}^j D^n \varphi^j)
-
(D_m \varphi^j D^m \bar{\varphi}^j)^2
\right]
- \frac{1}{4 \Lambda (X)}.
\end{align}
The equation of motion for $A_m$ is the same with \eqref{eq:cpn_nc} and it is again
solved by \eqref{eq:cpn_nc_gauge_sol}. Substituting this into the Lagrangian, 
the gauge covariant derivative $D_m \varphi^i$ is replaced by $\tilde{D}_m \varphi^i$.
The remaining constraints \eqref{eq:Qc0} and \eqref{eq:Qc1} are solved by the parametrizations 
\eqref{eq:Qphi1} and \eqref{eq:Qphi2}
and the target space is $Q^{N-2}$.
However, we should keep in mind that there is an extra constraint \eqref{eq:Qc2} given by
\begin{align}
\varphi^1 F^N (\tilde{X}) + \varphi^N F^1 (\tilde{X}) + \sum_{\hat{i}=2}^{N-1} \varphi^{\hat{i}} F^{\hat{i}} (\tilde{X}) = 0, 
\label{eq:Qn_F_constraint}
\end{align}
where $F^i (\tilde{X}) \not= 0$ is the solution to the
auxiliary field in the non-canonical branch.
The constraint \eqref{eq:Qn_F_constraint} generically contains space-time derivatives of the
fields $\varphi^i, \bar{\varphi}^{\bar{i}}$ and this has to be solved together with the following equation of motion:
\begin{align}
&
\Lambda'(\tilde{X}) \tilde{D}^m \varphi^i 
\left[
(\tilde{D}_m \bar{\varphi}^j \tilde{D}_n \varphi^j) 
(\tilde{D}^m \bar{\varphi}^k \tilde{D}^n \varphi^k)
-
(\tilde{D}_m \varphi^j \tilde{D}^m \bar{\varphi}^j)^2
\right]
\notag \\
&
- 2 \Lambda (\tilde{X})
\left[
\tilde{D}_n \varphi^i (\tilde{D}^m \bar{\varphi}^j \tilde{D}^n
 \varphi^j)
- \tilde{D}_m \varphi^i (\tilde{D}_n \varphi^j \tilde{D}^n \bar{\varphi}^j)
\right]
+ \frac{1}{4 \Lambda^2 (\tilde{X})} \Lambda'(\tilde{X}) \tilde{D}^m
 \varphi^i = 0,
\label{eq:QNC_eom}
\end{align}
where the prime in $\Lambda'(\tilde{X})$ denotes the differentiation with
respect to $\tilde{X}$.
In general, it is uncertain whether the simultaneous equations \eqref{eq:Qn_F_constraint} and \eqref{eq:QNC_eom}
admit non-trivial solutions or not.
Instead, the most plausible way to work with these equations is to restrict the dynamics to a subspace in $Q^{N-2}$.
This subspace highly depends on explicit structures of the solution
$F^i$.
For example, if we consider the $N=4$ case, 
one finds that a solution to the equation of motion for $F^i$ in the
non-canonical branch is
\begin{align}
F^2 = F^3 = F^4 = 0, \qquad F^1 = 
\sqrt{
- \frac{1}{2 \Lambda (\tilde{X})} + \tilde{D}_m \varphi^i \tilde{D}^m \bar{\varphi}^i
}.
\end{align}
Substituting this solution to the constraint \eqref{eq:Qn_F_constraint}
gives a condition 
\begin{align}
\varphi^4 \sqrt{
- \frac{1}{2 \Lambda (\tilde{X})} + \tilde{D}_m \varphi^i \tilde{D}^m \bar{\varphi}^i
} = 0.
\label{eq:Q1_constraint_example}
\end{align}
In order that this is compatible with equation of motion, we first
choose the subspace $\varphi^4 = 0$ in $Q^{2}$, and then solve the equation
\eqref{eq:QNC_eom} in that subspace.

A comment is in order.
As for the Grassmann case, we have an equality $Q^{4} = G_{4,2}$.
The constraint \eqref{eq:Qn_F_constraint} is necessary in 
the construction of the $Q^4$ sigma model discussed in this section but such a
restriction to a subspace was absent in the construction of the
$G_{4,2}$ model in the previous section.
Apparently, they have different Lagrangians even though 
there is a target space isomorphism between them.
The same is true for the relation $Q^1 = G_{2,1} =
\mathbb{C}P^1$.
In the non-canonical branch, 
the different constructions of sigma models result in the different
derivative corrections 
and the field space that the dynamics occurs, 
although they give identical models in the canonical branches.

\subsection{$SO(2N)/U(N)$ and $Sp(N)/U(N)$ models}
Observed that the $Q^{N-2}$ model has remarkable structures in the
non-canonical branch due to the F-term constraint, 
we next consider nonlinear sigma models
with the target spaces $SO(2N)/U(N)$ or $Sp(N)/U(N)$.
The gauged linear sigma model 
has the global symmetry 
$G = SO(2N)$ or $Sp(N) [=USp(2N)]$
and gauge symmetry $G'=U(N)$.
The chiral superfield $\Phi$ belongs to the $(\mathbf{2N}, \bar{\mathbf{N}})$
representation of $G \times G'$ and it is expressed as an $2N \times N$ matrix.
The constraint is given by the F-term superpotential
\begin{align}
W = \mathrm{Tr} 
\left[
\Phi_0 \Phi^t J' \Phi
\right],
\label{eq:SOSp_superpotential}
\end{align}
where $\Phi_0$ is an $N \times N$ symmetric (anti-symmetric) matrix
superfield satisfying $\Phi_0^t = \epsilon \Phi_0$ 
for $G=SO(2N) [Sp(N)]$.
Here $\epsilon = 1$ for $G = SO(2N)$ and $\epsilon = -1$ for $G= Sp(N)$.
The $U(1)_D \subset U(N)$ charge of $\Phi_0$ is assigned to $-2$ to
cancel the $U(1)$ charge $+1$ of $\Phi$.
The $2N \times 2N$ matrix $J'$ is given by
\begin{align}
J' = 
\left(
\begin{array}{cc}
0 & \mathbf{1}_N \\
\epsilon \mathbf{1}_N & 0
\end{array}
\right),
\end{align}
which is an invariant tensor of $SO(2N)$ ($\epsilon = 1$) or $Sp(N)$
($\epsilon = -1$).

The equation of motion for $D$ gives the constraint \eqref{eq:D-constraint2} 
while that for $F_0$ gives the constraint
\begin{align}
\varphi^t J' \varphi = 0,
\label{eq:SOSp_pJp}
\end{align}
and that for $\varphi_0$ gives a constraint
\begin{align}
\varphi^t J' F = 0.
\label{eq:SOSp_pJp2}
\end{align}
The equation for $F$ is \eqref{eq:eqF3}, but the gauge symmetry is non-Abelian.
One finds that $\varphi_0 = 0$ and $F = 0$ are a solution to the equation
of motion for $F$. 
Therefore the constraint \eqref{eq:SOSp_pJp2} becomes trivial in the canonical branch.
We find that the gauge field is integrated out by the solution 
$A_m = i \kappa^{-1} \bar{\varphi} \partial_m \varphi$ in the sigma
model limit which is the same with the case without the derivative corrections.
Then the target space of the sigma model is nothing but the
$SO(2N)/U(N)$ or $Sp(N)/U(N)$ and the 
derivative corrections appear as in the 
Grassmann case in Eq.~\eqref{eq:dc_grassmann}.

Compared with the $Q^{N-2}$ model, 
it is not straightforward to write down solutions $F \not= 0$ explicitly
due to the non-Abelian gauge symmetry.
However, similar to the Grassmann case, the solution
$\mathrm{Tr} [F\bar{F}]$ is given by the gauge invariant quantities 
such as $\mathrm{Tr} [D_m \varphi D^m \bar{\varphi}]$.
Therefore we again find that the gauge field $A_m$ is integrated out by
$A_m = i \kappa^{-1} \bar{\varphi} \partial_m \varphi$.
The subsequent discussion is parallel to the $Q^{N-2}$ case.
Due to the extra constraint \eqref{eq:SOSp_pJp2}, we find that 
the target spaces $SO(2N)/U(N)$ or $Sp(N)/U(N)$ should be restricted to their
subspaces in the non-canonical branches.

\subsection{$E_6/[SO(10) \times U(1)]$ and $E_7/[E_6 \times U(1)]$ models}

For the case of the target space $E_6/[SO(10) \times U(1)]$, 
the global symmetry is $G=E_6$ and the $G'=U(1)$ symmetry is gauged 
in the gauged linear sigma model \cite{Higashijima:1999ki}.
The chiral superfield $\Phi$ belongs to the fundamental representation
$\mathbf{27}$ of $E_6$.
This is decomposed into the maximal subgroup $SO(10) \times U(1)$.
The F-term constraint is given by the superpotential
\begin{align}
W = \Gamma_{ijk} \Phi_0^i \Phi^j \Phi^k,
\end{align}
where $\Gamma_{ijk}$ is a rank-3 symmetric invariant tensor of $E_6$.
Since the gauge symmetry is Abelian, the structure of the sigma model is
essentially the same with the $Q^{N-2}$ case.
Due to the constraint 
\begin{align}
\Gamma_{ijk} \varphi^i F^k = 0
\end{align}
derived by the superpotential, the target space of the sigma model is
restricted to a subspace of $E_6/[SO(10) \times U(1)]$ in the
non-canonical branch.

\medskip
For the target space $E_7/[E_6 \times U(1)]$, the global symmetry is
$G = E_7$ and the $G'=U(1)$ symmetry is gauged 
in the gauge linear sigma model.
The F-term constraint is \cite{Higashijima:1999ki}
\begin{align}
W = d_{\alpha \beta \gamma \delta} \Phi_0^{\alpha} \Phi^{\beta}
 \Phi^{\gamma} \Phi^{\delta},
\label{eq:E7_superpotential}
\end{align}
where $d_{\alpha \beta \gamma \delta}$ is a rank-4 symmetric invariant
tensor of $E_7$.
Since the gauge symmetry is Abelian, the structure of the sigma model is
essentially the same with the $Q^{N-2}$ case.
Again, the constraint 
\begin{align}
d_{\alpha \beta \gamma \delta} \varphi^{\alpha} \varphi^{\beta}
 F^{\gamma} = 0
\end{align}
defines a subspace in $E_7/[E_6 \times U(1)]$ in the non-canonical branch.

\section{Bogomol'nyi-Prasad-Sommerfield states} 
\label{sect:BPS}
In this section, we study the BPS conditions in the nonlinear sigma models discussed in
the previous section. 
BPS properties of Skyrme type and higher derivative models have been studied in various
contexts \cite{Adam:2018pvd, Casana:2019zlb, Casana:2019ztm, Stepienn:2019fdk}.
It has been discussed that the gauged chiral models admit 
the 1/2 BPS vortex state and the 1/4 BPS states \cite{Nitta:2015uba}.
In the following, we derive the 1/2 and 1/4 BPS conditions in the
nonlinear sigma models in the limit $g \to \infty$ of those in the gauged chiral models.

\subsection{BPS states in canonical branch}

In the canonical branch, the 1/2 BPS vortex equation for the finite $g$ 
for gauged linear sigma model 
is 
\begin{align}
\bar{D}_z \varphi^i = 0, \qquad \frac{1}{g} F_{12} - (\bar{\varphi} \varphi -
 \kappa \mathbf{1}) = 0,
\label{eq:half_BPS}
\end{align}
where $z = x^1 + i x^2$ is the complex coordinate in the
$(x^1,x^2)$-plane.
These are $U(M)$ BPS semilocal vortex equations with $N$-flavors
in the case of Grassmann case \cite{Shifman:2006kd,Hanany:2003hp},
for which we have found that higher derivative corrections do not appear.
They reduce to the ordinary BPS vortex equations in the
Abelian-Higgs model for the $U(1)$ gauge theory.

The energy bound for this configuration is \cite{Nitta:2015uba}\footnote{
We comment on the positive-semi definiteness of the energy in the higher
derivative chiral models. 
It is not always true that the energy density derived from
Eq.~\eqref{eq:Lagrangian} is positive-semi definite for arbitrary $\Lambda$
and the K\"ahler potential $K$.
In order to define the BPS states as minima of energy, we have to choose
appropriate $\Lambda$ and $K$ that makes the energy be positive-semi definite \cite{Nitta:2014pwa}.
}
\begin{align}
\mathcal{E} = - \kappa F^0_{12}.
\label{eq:energy_bound}
\end{align}
The configuration is classified by the vortex number $\int \! d^2 x \, F^0_{12}$.

In the nonlinear sigma model limit $g \to \infty$, the second condition in \eqref{eq:half_BPS} gives the D-term
constraint while the first one becomes
\begin{align}
\bar{\partial}_z \varphi^i - \kappa^{-1} (\bar{\varphi}^j
 \bar{\partial}_z \varphi^j) \varphi^i = 0.
\end{align}
This implies the following 1/2 BPS lump equation in the sigma models:
\begin{align}
\bar{\partial}_z \varphi^i = 0.
\label{eq:cb_nlsm_BPS}
\end{align}
The bound \eqref{eq:energy_bound} survives in the limit $g \to \infty$ and the result is 
\begin{align}
\mathcal{E} =  \frac{1}{2} |\partial_z \varphi|^2.
\label{eq:cb_nlsm_BPS_bound}
\end{align}
This provides the lump charge density in the nonlinear sigma model.
We note that the derivative corrections never show up in the equation
\eqref{eq:cb_nlsm_BPS} and the BPS bound \eqref{eq:cb_nlsm_BPS_bound} in
the canonical branch.\footnote{
As denoted below Eq.~(\ref{eq:CPN}),  
for the case of the ${\mathbb C}P^{N-1}$ model, 
the Lagrangian in Eq.~(\ref{eq:CPN}) (with $\Lambda=$const.)
can be obtained 
\cite{Eto:2012qda}
as the low-energy effective theory of 
a BPS non-Abelian vortex in ${\cal N}=2$ supersymmetric $U(N)$ gauge theory 
\cite{Hanany:2003hp}.
In this context, 
${\mathbb C}P^{N-1}$ lumps (sigma model instantons) 
on the vortex worldsheet 
represent Yang-Mills instantons in the bulk theory,
and such instanton-vortex composites are 1/4 BPS states
\cite{Hanany:2004ea}.
It was discussed in Ref.~\cite{Eto:2012qda} that 
this fact implies that lump solutions should not have derivative corrections.
}

\subsection{BPS states in non-canonical branch}
In the non-canonical branch, we have only 1/4 BPS state \cite{Nitta:2015uba}.
The BPS equations in the gauged linear sigma models are
\begin{align}
\bar{D}_z \varphi = - i F, \qquad \frac{1}{g} F_{12} = \bar{\varphi}
 \varphi -
 \kappa \mathbf{1}.
\end{align}

In the nonlinear sigma model limit $g \to \infty$, the BPS equations become
\begin{align}
\bar{D}_z \varphi = - i F, \qquad \bar{\varphi} \varphi = \kappa \mathbf{1}.
\end{align}
The latter equation gives the D-term constraint.
From the first equation, we obtain
\begin{align}
\bar{D}_z \varphi D_z \bar{\varphi} = F \bar{F}.
\end{align}
For the $\mathbb{C}P^{N-1}$, $Q^{N-2}$ cases, the gauge group is Abelian and
we find an explicit solution for $F$. Then the BPS condition is
rewritten as 
\begin{align}
\bar{\tilde{D}}_z \bar{\varphi} \tilde{D}_z \varphi -
 \bar{\tilde{D}}_z \varphi^i \tilde{D}_z \bar{\varphi}^i = \Lambda^{-1}
 (\varphi, \bar{\varphi}).
\label{eq:nc_BPS}
\end{align}
This is nothing but a compacton type equation.
To see this, we choose an appropriate function $\Lambda$.
For example, we consider the following function in the $\mathbb{C}P^1$ model,
\begin{align}
\Lambda = 2 
\left(
\frac{|u|^2}{1+|u|^2}
\right)^{- \frac{s}{2}},
\end{align}
where $s$ is a constant.
Then, the BPS equation \eqref{eq:nc_BPS} results
in the following compacton equation \cite{Adam:2009px}:
\begin{align}
n g_y = - g^{\frac{s}{2}}.
\end{align}
Here we have assumed the ansatz
\begin{align}
u = e^{in\theta} f(r), \quad 
1 - g = \frac{1}{1+f^2}, \quad y = \frac{r^2}{2},
\end{align}
where $r$ and $\theta$ are the polar coordinates in the $(x^1,x^2)$-plane.

\section{Comment on fermionic interactions}\label{sect:fermion}
Finally, we comment on the fermionic interactions of the models,
while in the previous sections, we have focused on the bosonic sector of the supersymmetric models and
have shown that the vector field $A_m$ is integrated out exactly.
However, the integration of the gaugino is rather awkward 
from the reason described below.
We illustrate the problem in the $\mathbb{C}P^{N-1}$ model.

Assuming that $\Lambda$ does not contain $V$ for simplicity,
the equation of motion for the vector superfield $V$ in the sigma model
limit $g \to \infty$ is given by 
\begin{align}
& \quad 
2 \Phi^i e^{2V} \Phi^{\dagger i} - \kappa 
\notag \\
&  
+ \frac{1}{4} \Lambda (\Phi, \Phi^{\dagger}) e^{4V} 
\mathcal{D}^{\alpha} \Phi^i \mathcal{D}_{\alpha} \Phi^j
 \bar{\mathcal{D}}_{\dot{\alpha}} \Phi^{\dagger i}
 \bar{\mathcal{D}}^{\dot{\alpha}} \Phi^{\dagger j} 
\notag \\
&  
+ \frac{1}{4} D_{\alpha} 
\left[
\Lambda (\Phi, \Phi^{\dagger}) e^{4V} \Phi^i \mathcal{D}^{\alpha} \Phi^j
\bar{\mathcal{D}}_{\dot{\alpha}} \Phi^{\dagger i}
 \bar{\mathcal{D}}^{\dot{\alpha}} \Phi^{\dagger j}
\right]
\notag \\
&  
+ \frac{1}{4} \bar{D}_{\dot{\alpha}} 
\left[
\Lambda (\Phi, \Phi^{\dagger}) e^{4V} 
\mathcal{D}_{\alpha} \Phi^i \mathcal{D}^{\alpha} \Phi^j \Phi^{\dagger i}
 \bar{\mathcal{D}}^{\dot{\alpha}} \Phi^{\dagger j}
\right] = 0.
\label{eq:Veom}
\end{align}
If the term in the second line is ignored, we find that the vector superfield
is solved by 
\begin{align}
V = - \frac{1}{2}\log (2 \kappa^{-1} \Phi^i \Phi^{\dagger i}).
\label{eq:Vsol}
\end{align}
This is easily confirmed if one notices that the following relation holds for 
Eq.~\eqref{eq:Vsol}:
\begin{align}
\Phi^{\dagger i} \mathcal{D}_{\alpha} \Phi^i = 
\Phi^{\dagger i}
\left[
D_{\alpha} \Phi^i + 2 (D_{\alpha} V) \Phi^i 
\right]
= \Phi^{\dagger i} D_{\alpha} \Phi^i - \frac{1}{|\Phi|^2} \Phi^{\dagger j}
 D_{\alpha} \Phi^j ( \Phi^i \Phi^{\dagger i})
= 0.
\end{align}
The would-be solution \eqref{eq:Vsol},
whose $\theta \sigma^m \bar{\theta}$ component gives $A_m = i \kappa^{-1} \bar{\varphi}^i \del_m \varphi^i$,
precisely gives the correct K\"ahler
potential $K = - \log (2\kappa^{-1} \Phi \Phi^{\dagger})$ for the 
$\mathbb{C}P^{N-1}$ model.
However, when the term in the second line in Eq.~\eqref{eq:Veom}
is included, Eq.~\eqref{eq:Vsol} fails to be a solution.
This implies that the expression in Eq.~\eqref{eq:Vsol} is correct only in the bosonic sector.
Including the second line in Eq.~\eqref{eq:Veom} modifies the equality in Eq.~\eqref{eq:Vsol}
in the fermionic sector.

We note that in solving the auxiliary field $F$, the fermionic fields
$\psi$ in the chiral multiplets are introduced 
perturbatively around the bosonic solution $F = F(\varphi,
\bar{\varphi})$ \cite{Khoury:2010gb}.
The same is true for the vector multiplet.
Since the interaction terms are all expressed by the superfields
explicitly, it is in principle possible to write down the equations for
the fermions. 
We can integrate out the vector multiplet starting from the bosonic part of the solution 
\eqref{eq:Vsol} and introduce the fermions perturbatively.
This procedure determines the fermionic interactions in the sigma model
limit $g \to \infty$.
Even though the fermionic interactions are rather cumbersome, 
we stress that the BPS conditions in the nonlinear sigma models are 
not affected by details of the fermionic interactions. 
As we have shown, this is obtained by those in the gauged chiral models.

\section{Conclusion and discussions} \label{sect:conclusion}
In this paper, we have constructed $\mathcal{N} = 1$ supersymmetric higher derivative 
terms in 
nonlinear sigma models whose target spaces are Hermitian symmetric
spaces.
We have considered the sigma model limit of the supersymmetric
gauged linear sigma models involving the derivative terms for chiral
multiplets that are free from the auxiliary field problem 
and the Ostrogradski's ghost.
In the sigma model limit, the vector multiplet does not propagate anymore and
can be integrated out.
Due to the supersymmetric derivative corrections, there are two distinct
on-shell branches called the canonical and non-canonical branches.
This structure is carried over to the nonlinear sigma models.
We have shown that the gauge field $A_m$ is explicitly integrated out,
both in the canonical and the non-canonical branches, even in the presence of 
higher 
derivative terms.

In the canonical branch, where the auxiliary field is given by $F = 0$, 
we explicitly have written down the Lagrangians of the nonlinear sigma models.
They consist of the canonical kinetic terms and the derivative
corrections characterized by the arbitrary function $\Lambda$.
This is a natural generalization of the sigma model construction
discussed in Ref.~\cite{Higashijima:1999ki}.  

For the $\mathbb{C}P^{N-1}$ model, we have obtained the supersymmetric 
$\mathbb{C}P^{N-1}$ Skyrme-Faddeev model. 
In the non-canonical branch, where the auxiliary field is given by $F
\not= 0$, the situation changes drastically.
For the $\mathbb{C}P^{N-1}$ model, an Abelian symmetry is gauged and we have 
found the explicit solutions of the non-zero auxiliary field $F$.
We have solved all the constraints and 
have written down the explicit Lagrangians of the $\mathbb{C}P^{N-1}$
nonlinear sigma model.
The canonical quadratic kinetic term is absent in the non-canonical
branch, and there is only the fourth derivative term
as a generalization of the Skyrme-Faddeev term of 
the $\mathbb{C}P^1$ model.
We thus have obtained the supersymmetric $\mathbb{C}P^{N-1}$ Skyrme-Faddeev model.

For the formulation of the $G_{M,N}$ model, a non-Abelian $U(N)$ symmetry is
gauged in the gauged linear sigma model. 
Compared with the $\mathbb{C}P^{N-1}$ case for which 
the gauge symmetry is
Abelian in the gauged linear sigma model, the equation of motion for the auxiliary field in 
the $G_{M,N}$ model is rather complicated.
Although it is hard to find explicit $F\not=0$ solutions, we have been able to 
integrate out
the gauge field with the help of the gauge invariance of the higher derivative
terms.

For other Hermitian symmetric spaces, 
in addition to D-term constraints, 
we further impose F-term constraints yielding holomorphic 
embedding of the target spaces into  
 $\mathbb{C}P^{N-1}$ or Grassmann manifold, 
 as was done for the case without higher derivative terms.
 In the canonical branches, these constraints are consistent 
 in the presence of higher derivative terms, 
 but we find that 
 in the non-canonical branch these constraints 
 yield further additional constraints reducing the 
 target spaces to their submanifolds.

For the formulation of the $Q^{N-2}$ model, an Abelian symmetry is gauged 
in the gauged linear sigma model, 
but we have an extra constraint coming from the F-term 
embedding $Q^{N-2}$ to  $\mathbb{C}P^{N-1}$. 
The non-zero solution $F \not= 0$ yields
an extra constraint on the scalar fields.
We have found that in order that 
these constraints are compatible with the equation of motion
in the non-canonical branches,
the dynamics of fields should be restricted to a subspace in
the $Q^{N-2}$ manifold, 
which is a peculiar property of this model.
We have also discussed the sigma models whose target spaces are
$SO(2N)/U(N)$, $Sp(N)/U(N)$, $E_6/[SO(10) \times U(1)]$ and $E_7/[E_6
\times U(1)]$.
The first 
two cases are
essentially similar to the cases of the Grassmann and the
$Q^{N-2}$ cases.
For the latter two cases, 
although the superpotential is complicated compared with the $Q^{N-2}$
case, the structure of the higher derivative terms is similar to that of the $Q^{N-2}$ case.
We also have found that these constructions provide the different sigma
models even though there are several isomorphisms among the target spaces
such as $G_{N,1} \simeq G_{N,N-1} \simeq \mathbb{C}P^{N-1}$,
$Q^{1} \simeq G_{2,1} \simeq \mathbb{C}P^1$, $Q^4 \simeq G_{4,2}$
and so on.

Finally, we provide a comment on the BPS equations and the integration of the
fermionic terms in the vector multiplet.
In the canonical branches,  
BPS equations and their solutions 
are the same 
with those without higher derivative terms, 
while, in the non-canonical branches they give compacton type 
configurations.

It would be interesting to study explicit solutions to the BPS
equations in the non-canonical branch.
As we have noted, the 1/4 BPS equation in the non-canonical branch of
the $\mathbb{C}P^{N-1}$ model reduces to that of the compactons.
Finding solutions in the other sigma models is interesting.
We will come back to these issues in future studies.

In this paper, we have considered the case that there are no potential terms in 
the nonlinear sigma models (although we have introduced superpotentials 
for the F-term constraints 
in the gauged linear sigma models).
If we also introduce superpotentials or twisted masses (from dimensional reductions), 
there are more BPS states such as domain walls, their junctions and so on 
\cite{Nitta:2014pwa}.
Introducing potential terms and studying associated BPS states remain as a future work.

\subsection*{Acknowledgments}
This work was supported in part by Grant-in-Aid for Scientific
Research, JSPS KAKENHI Grant Numbers 
JP18H01217 (M.~N.) and 
JP20K03952 (S.~S.).


\end{document}